\documentclass[prb,onecolumn,nofootinbib,citeautoscript,10pt,longbibliography,notitlepage]{revtex4-2}

\pdfoutput=1

\usepackage{graphicx}
\usepackage{dcolumn}
\usepackage{color}
\usepackage{amssymb,amsmath}
\usepackage{tabularx,graphicx}
\usepackage{epstopdf}
\usepackage{latexsym}
\usepackage{colortbl}
\usepackage{psfrag}
\usepackage{bbm,bm,array,physics}
\usepackage{titlesec}
\usepackage{dsfont}
\usepackage[tight]{subfigure}

\usepackage[papersize={8.5in,11in}]{geometry}

\definecolor{darkblue}{rgb}{0.,0.,0.4}
\definecolor{darkred}{rgb}{0.5,0.,0.}
\definecolor{BlueViolet}{RGB}{138,43,226}
\definecolor{SkyBlue}{RGB}{30,144,255}
\definecolor{DarkGreen}{RGB}{0,100,0}
\usepackage[pdftex,colorlinks=true,linkcolor=darkblue,
citecolor=blue,urlcolor=darkred]{hyperref}
\def\*#1{\mathbf{#1}} 

\begin{document}
\title{Signatures of two- and three-dimensional semimetals from circular dichroism}

\author{Ipsita Mandal}
\affiliation{Institute of Nuclear Physics, Polish Academy of Sciences, 31-342 Krak\'{o}w, Poland}

\begin{abstract} 
Topological invariants are crucial quantities for classifying materials with topological phases. Hence, their connections with experimentally measurable quantities are extremely important. In this context, circular dichroism (CD) provides a protocol to detect the Chern number $\mathcal{C}_0$ of the lowest energy Bloch band (LBB) of a semimetal. This hinges on the unequal depletion rates of the Bloch electrons from a filled LBB, under the action of a time-periodic circular drive, depending on the chirality of the polarization. According to the dimensionality of the system (i.e., whether it is two- or three-dimensional), the integrated differential rate for depletion has to be formulated a bit differently in order to relate it to $\mathcal{C}_0$. Our aim is to capture the nature of the CD response for semimetals with anisotropic band dispersions. We show that while the quantization of the CD response for the three-dimensional cases is strongly sensitive to anisotropy, the two-dimensional counterparts show a perfectly quantized response.
\end{abstract}

\maketitle


\section{Introduction}
\label{sec:intro}

A Berry phase is a geometrical phase angle that describes how a global phase accumulates over the course of a cycle, as some complex vector is transported around a closed loop in a complex vector space \cite{pancharatnam,berry,wilczek}. Berry phases manifest themselves
in a rich variety of physical contexts including classical optics, atomic and molecular physics, nuclear physics,
photonics, and condensed-matter physics. The Berry connection is the local gauge field (analogous to the vector potential) associated with the Berry phase, whose exterior derivative gives us the Berry curvature (BC). The BC turns out to be the imaginary part of the quantum geometric tensor \cite{vanderbilt}. The integral of the BC over a closed manifold is quantized in units of $2\pi $, and is referred to as the Chern number. 

In solid state physics, the Berry phase enters into the quantum-mechanical band theory of electrons in crystals, and the complex vector in question is a Bloch wavevector, such that the closed path lies in the space of wavevectors within the Brillouin zone. As a result, the Chern number is defined entirely in terms of the momentum space wave functions, and has emerged as an important tool to classify topological phases of matter.

Over the past few years, the impact of BC on various transport phenomena
has attracted tremendous interest. Some examples include anomalous Hall effect~\cite{nagosa_anomalous}, Magnus Hall effect~\cite{papaj_magnus,amit-magnus,sajid_magnus}, negative magnetoresistance~\cite{li_nmr17,dai_nmr}, nonlinear Hall effect~\cite{he_nonlinear}, planar Hall effect \cite{Nandy_2017,Nag_2020,shivam-serena}, circular photogalvanic effect~\cite{juan_quantized,kozii,Mandal_2020}, circular
dichroism~\cite{kush-cd,goldman_dir,goldman20,sajid_cd}, and second-order photoconductivities in terms of geometrical quantities~\cite{nonlin-photo}.

\begin{figure}[]
	\centering
	\includegraphics[width= 0.35 \textwidth]{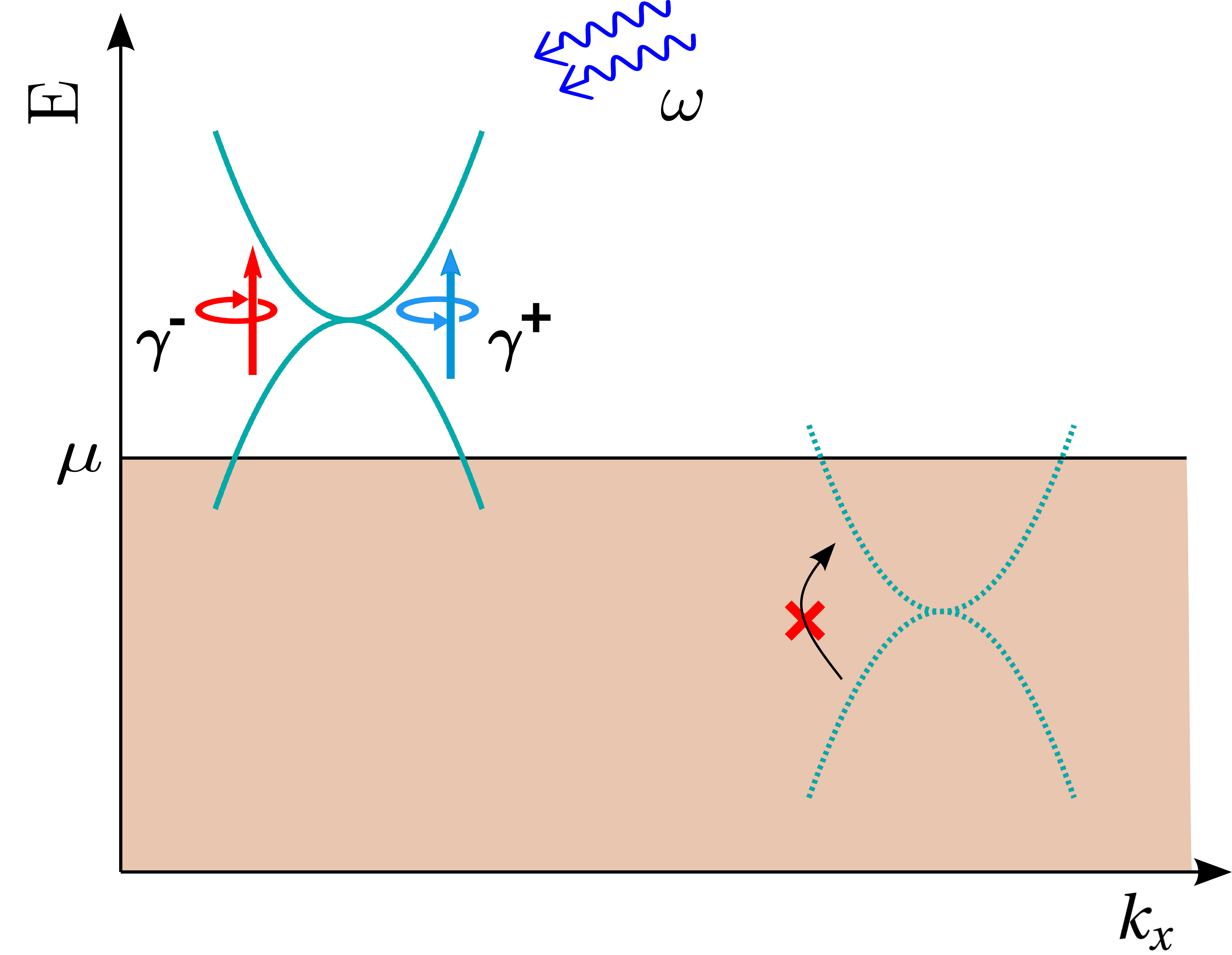}
	\caption{\label{fig:set-up}
Schematics representing the central idea behind the CD response: The solid and dashed cyan curves show dispersion $E$ of a double-Weyl semimetal against the $k_x$-component of the momentum vector, with $\mu$ denoting the Fermi level. Absorption of incident photons of energy quantum $ \hbar\, \omega$ leads to chiral transition rates $\gamma^{\pm}$, depending on the chirality of the external drive. Since we consider nodes of opposite chiralities ($\pm 1 $) lying at different energies, the Fermi level has been tuned to Pauli-block the transitions within the double-Weyl node on the right-hand-side (shown by dashed cyan curves). In such a situation, only the transitions from the left-hand-side double-Weyl node (shown by solid cyan curves) contribute to the CD response.}
\end{figure}

In recent times, several techniques~\cite{dai,stefanos,turner} across diverse platforms have been proposed to detect the Chern numbers of the Bloch bands. Among them, ``circular dichroism'' (CD) is a response under a circular time-periodic perturbation~\cite{goldman_dir,goldman20,asteria} (for example, by shining circularly polarized light). The phenomenon hinges on the fact that the depletion rate of the filled bands in a topological phase, under the effects of an external drive, depends on the orientation (handedness) of the circular shake. The chiral nature of the system causes it to have unequal depletion rates depending on whether the drive is left- or
right-handed.\footnote{This chirality is in analogy with the chiral Majorana quasiparticles that appear at the edges of topological nanowires~\cite{kitaev,ips-sudip,ips-tewari,ips-epl,ips-counting,ips-jay}. In the context of the semimetals, the sign of the topological charge (or in turn, the chirality) can be related to the chiral edge states observed in various experiments \cite{Howard2021}.} This central idea behind the CD response is depicted schematically in Fig.~\ref{fig:set-up}.
Circular drives, confined to a two-dimensionsional (2d) plane, can be realized through circular shaking in ultracold atoms trapped in optical lattices~\cite{eckardt}, using piezo-electric actuators~\cite{jotzu} --- this activates transitions involving momentum components (of the Bloch states) confined to that particular 2d plane. In fact, this technique has been implemented to measure the Chern numbers of systems like the two-band Haldane model~\cite{goldman_dir} and topological floquet bands~\cite{asteria}. Furthermore, a similar technique with linear shaking has been proposed~\cite{ozawa} to probe the quantum metric tensor in periodically driven systems. 

In 2d systems, the net depletion rate integrated over the momenta and frequency domain yields a quantized observable dubbed as the ``differential integrated rate'' (DIR), and is proportional to the Chern number $\mathcal{C}_0$ of the lowest energy Bloch band~\cite{goldman_dir}. The quantized response has been demonstrated in a variety of systems like 2d topological insulators, transition metal dichalcogenides~\cite{shan22}, 2d quantum magnets~\cite{sentef23}, as well as amorphous systems~\cite{marsal20}.

In our earlier paper~\cite{sajid_cd}, we have formulated the CD response for three-dimensional (3d) systems. The prescription involves computing the product of the net depletion rate and the difference of band velocities, which is then integrated over the 3d Brillouin zone. We call this quantity the 3d DIR, which is found to be quantized for topological semimetals having isotropic dispersions. However, the quantization is found to be strongly affected if anisotropies are included. With this backdrop, we compute the CD responses for both 2d and 3d anisotropic semimetals. Specifically, we consider multi-Weyl~\cite{bernevig_mwsm} and semi-Dirac~\cite{pardo09, pardo_prb} semimetals that feature anisotropic dispersions. Although band-touching nodes always come in pairs due to the Nielsen-Ninomiya theorem~\cite{nielsen}, with each pair featuring opposite chiralities, here we consider a single isolated node for computing the CD. This is because if the inversion and mirror symmetries are broken, conjugate nodes with opposite chiralities appear at different energies~\cite{burkov}, in which case we can Pauli-block the one at lower energies by adjusting the Fermi level (cf. Fig.~\ref{fig:set-up}).

The paper is organized as follows. In Sec.~\ref{sec:formalism}, we elucidate the formalism to obtain the 2d and 3d DIR response. We then numerically compute the response for various systems in Sec.~\ref{sec:results}. We end with a summary and outlook in Sec.~\ref{sec:summary}.

\section{Circular drive and differential integrated rates in 2d and 3d}
\label{sec:formalism}

Let us consider a circularly polarized time-periodic drive described by the Hamiltonian $\mathcal{H}_{\pm}^{\prime}(t) = 2\, \mathcal{E} \left [-\sin(\omega\, t) \, \hat{\mu}\pm\cos(\omega \, t) \, \hat{\nu} \right]$, where $\mathcal E$ is the amplitude, $\omega$ is the frequency, $\hat \mu $ and $\hat \nu$ denote the mutually
perpendicular position vector operators defining the plane of polarization of the applied drive, and the subscript ``$\pm$'' refers to the handedness of the circular polarization.
A non-interacting Hamiltonian $\mathcal H_0(\*k)$ (describing the semimetal in question) is subjected to this perturbation, such that the total Hamiltonian is given by
\begin{align}
 \mathcal{H}_{\pm}^{\text{tot}}(\*k,t)= \mathcal H_0(\*k) + \mathcal{H}_{\pm}^{\prime}(t)\,.
\end{align}
It is convenient to use a unitary transformation to bring the Hamiltonian to a translation-invariant form~\cite{goldman_dir,ozawa}. In addition, we assume small driving amplitudes $\mathcal{E}/(\hbar \,\omega) \ll 1$ [where $\hbar=h/(2\pi)$ is the reduced Planck's constant], because the drive is treated as a perturbation. Incorporating these ingredients, the final Hamiltonian takes the form:
\begin{align}
{\tilde{\mathcal H}}_{\pm}^{\text{tot}}	(\*k, t) \approx \mathcal{H}_0(\*k) + \frac{2\, \mathcal{E}}{\hbar\, \omega} \, 
	\left[   \cos(\omega \, t) \, \frac{\partial \mathcal{H}_0(\*k)}{\partial k_{\mu}} \pm \sin(\omega \, t) 
\, \frac{\partial \mathcal{H}_0(\*k)}{\partial k_{\nu}} \right ]  ,
\end{align} 
to leading order in $\mathcal{E}/\left( \hbar\, \omega \right)$.

For a two-band Hamiltonian, at the initial time $t=0$, we consider having a system with the lower energy Bloch band (with eigenvalue $ E_0(\*k)$ and eigenstate $\ket{0}$) filled. Then on switching on the periodic drive starting from $t=0$,
transitions of the Bloch electrons to the upper band (with eigenvalue $E_1(\*k)$ and eigenstate $\ket{1}$) are activated. This leads to a ``depletion rate'' of the occupied Bloch band --- a pivotal quantity of the CD protocol. Using the Fermi's golden rule~\cite{sakurai}, the depletion rate for sufficiently long time scales is given by
\begin{align}
\gamma^{\pm}_{10} (\*k,\omega) = 
\frac{2\pi}{\hbar} \,  \left(  \frac{\mathcal{E}}{\hbar \, \omega }  \right )^2
	\Big| \mathcal{P}^{\mu}_{10}(\*k) \mp i \, \mathcal{P}^{\nu}_{10}(\*k) \Big|^2 \, 
\delta\big ( E_{10}-\hbar \, \omega \big ) \, ,
\end{align}
where $\mathcal {P}^{\mu}_{10}(\*k) = \bra{1} \partial_{k_{\mu}} 
\mathcal H_0(\*k) \ket{0}$ is the optical matrix element. Here, $E_{10}=E_1(\*k)-E_0(\*k)$ represents the energy difference between the higher and the lower energy bands.
Henceforth, we abbreviate the lowest energy Bloch band as LBB.

Clearly, the presence of the delta function in our analytical expression forces the photon energy to precisely match with the band energy difference $\epsilon_{\*k}$. But in practice, the transitions of course take place within a short energy window. Thus we adopt the definition of the delta function as a limit of a Lorentzian: $\delta(x) \equiv (1/\pi) \, \lim_{\varepsilon \to 0} \; \varepsilon /\left (\varepsilon ^2+x^2 \right )$, and approximate this as a Lorentzian with a small width $2\varepsilon$ in our numerics.\footnote{Specifically, we have set $\varepsilon = 10^{-3}$ in obtaining all the plots.} The 2d differential integrated rate (or DIR) is obtained by integrating the net depletion rate over the $\*k$-space and frequency, and takes the form
\cite{goldman_dir, goldman20,asteria}:
\begin{align}
\label{eqdir1}
\Gamma^{\text{int}}_{\text{2d}} & = \int d\omega \, \frac{dk_x\,dk_y} {(2\pi)^2}
 \, \frac{\gamma^{+}_{10}(\*k,\omega) 
- \gamma^{-}_{10}(\*k,\omega)}{2} \\
\Rightarrow
\label{eqdir2}
\Gamma^{\text{int}}_{\text{2d}} & =  \left(  \frac{\mathcal{E}}{\hbar}  \right )^2 \mathcal{C}_0 \,.
\end{align}
The above expression shows that the 2d DIR is quantized in units of $ \left(   {\mathcal{E}}/{\hbar}  \right )^2$, being proportional to the LBB Chern number $\mathcal{C}_0$. 

In order to compute the 3d DIR~\cite{sajid_cd}, we need to consider the fact that the normal to the plane of the circular drive can be oriented along one of the three mutually perpendicular Cartesian coordinate axes.
Hence, we multiply the net depletion rate $\Delta\gamma^{\mu\nu}_{10} $ (for the polarization of the drive confined to the $\mu\nu$-plane) with the component of $\boldsymbol{v}_{10}$ in the perpendicular direction, where $ \boldsymbol v_{10} $ is the difference of the band velocities. Additionally, the integration is done over the 3d momnetum space only, which renders the integrated rate as a function of $\omega$ as shown below:
\begin{align}
\label{eqdir3d}
\Gamma^{\text{int}}_{\text{3d}} (\omega) 
& = \frac{1}{2}   \int 
\frac{dk_x\,dk_y\,dk_z} {(2\pi)^3} \sum_{\mu,\nu,\lambda}
\epsilon_{\lambda\mu\nu} \,  v_{10}^{\lambda} (\*k) \, 
\Delta\gamma^{\mu\nu}_{10} (\*k,\omega) \,, \nonumber \\
\text{where } 
v^{\lambda}_{10} (\*k) & = 
\frac{1}{\hbar}\frac{\partial E_{10}} {\partial k_{\lambda}} 
	\, , \text{ and }
\quad \Delta\gamma^{\mu\nu}_{10} = \frac{\gamma^{+}_{10} (k_{\mu},k_{\nu},\omega)
-\gamma^{-}_{10} (k_{\mu},k_{\nu},\omega)}{2} \,.
\end{align}
Here, the Levi-Civita symbol $\epsilon_{\lambda\mu\nu}$ denotes cyclic permutation. For semimetals having isotropic dispersions, the 3d DIR has been shown to take the simple form~\cite{sajid_cd}:
\begin{align}
\label{eqdir3d1}
\Gamma^{\text{int}}_{\text{3d}} (\omega) = 
\frac{(\mathcal{E}/\hbar)^2}{2\pi} \,  \mathcal{C}_0 \, .
\end{align}
Hence, $\Gamma^{\text{int}}_{\text{3d}} (\omega)$ is always quantized [in units of $ \left(   {\mathcal{E}}/{\hbar}  \right )^2 / (2\pi)$] for an isotropic two-band model.

For a generic multiband system (here more than two bands might be present), the 3d DIR formula for an isotropic system can be expressed as
\cite{sajid_cd}
\begin{align}
\Gamma^{\text{int}}_{\text{3d}} (\omega) = 
\frac{(\mathcal{E}/\hbar)^2}{2\pi} \, \left[   \mathcal{C}_0 - \frac{i}{2\pi} \int_S d\*S \cdot \sum_{m\ne0,1} \frac{\boldsymbol{\mathcal{P}}_{m0}
\times \boldsymbol{\mathcal{P}}_{0m}}{ E^2_{m0}}\right ]  ,
\quad E_{m0} =E_m-E_0\,,\quad
\mathcal {P}^{\mu}_{m {m'}}(\*k) = \bra{m} \partial_{k_{\mu}} 
\mathcal H_0(\*k) \ket{m'} ,
\end{align}
using the Stoke's theorem, with $d\mathbf S$ denoting an infinitesimal area vector of the surface integral over the surface $S$. Here, we have used the index $m$ to label all the bands in the system (with $m=1$ denoting the first higher energy band above the LBB, with the latter labelled as $m=0$), and have considered the situation when the LBB is filled at time $t=0$.
We note that the second term is relevant only in a system with more than two bands (as $m>2$ in the summation there), and hence it drops out for a two-band system [thus reducing to the expression in Eq.~\eqref{eqdir3d1}].

\section{Numerical results} 
\label{sec:results}

In this section, we compute the CD response for various 2d and 3d semimetals having two bands crossing at nodal points or closed nodal curves. The continuum Hamiltonians representing the semimetals are only approximations of actual lattice Hamiltonians in the low-energy limits. Hence, the momentum variables appearing in the integrals require a physical cutoff. In our numerics, we define an isotropic momentum cutoff $\Lambda$, upto which the continuum approximation remains to a reasonable approximation. For the frequency integrals, we have set the upper limit to $\hbar \,\omega= E_{10}$. In addition, we have chosen natural units (i.e., we have set $\hbar=k_B=c=1$) for simplicity. 

In all our examples, we consider a single node (and set its chirality to have the positive value), without any loss of generality, as it is a valid approximation for materials with well-resolved nodes in momentum space (e.g., in $\mathrm{Ta_3S_2}$~\cite{hasan16}). For the two-band models that we consider in this section, we can rewrite the Hamiltonian as $\mathcal{H}= \*d_{\*k} \cdot \boldsymbol{\sigma}$, where $\*d_{\*k}$
denotes the vector comprising the three coefficients appearing against the three Pauli matrices. using this form, the Berry curvature of the LBB is simply given by the formula
\begin{align}
	\Omega^{\mu \nu} (\*k) = - \frac{1}{2} \, \hat{\*d}_{\*k} \cdot  \left(  
	\partial_{k_\mu } \hat{\*d}_{\*k} \times \partial_{k_\nu} \hat{\*d}_{\*k}   \right ) , 
\quad \text{where} \quad \hat{\*d}_{\*k} = \frac{\*d_{\*k}}{|\*d_{\*k}|} \, .
\end{align}

\subsection{Semi-Dirac semimetal}
\label{sec:sd}

Semi-Dirac semimetals are 2d systems having a quadratic dispersion in one direction and a linear dispersion along the orthogonal direction~\cite{pardo09,pardo_prb,pickett09,saha_awf,ips-kush}. Promising candidates for such semi-Dirac systems include multilayered $\mathrm{ViO_2}$/$\mathrm{TiO_2}$ nanostructures~\cite{pardo09,pardo_prb,pickett09}, organic salts \cite{kobayashi,suzumura}, and deformed graphene \cite{hasegawa,orignac,montambaux1,montambaux2}. The low-energy continuum model Hamiltonian describing a semi-Dirac material is captured by~\cite{pickett09,montambaux1,montambaux2,saha_awf}
\begin{align}
\mathcal{H}_{\text{sd}}^{0} (\*k) =  \left(  \frac{\hbar^2 
 k_x^2} {2 \, m_x }
-\delta_0  \right ) \, \sigma_x + \hbar \, v_y\, k_y \, \sigma_y
\Rightarrow
\frac{ \mathcal{H}_{\text{sd}}^{0} (\*k) } {\hbar\,v_y\, k_0} = 
 \left[  \frac{  k_x^2/k_0^2 }  {2 \, \frac{m_x\,v_y} {\hbar \,k_0} }
- \frac{\delta_0} {\hbar\,v_y\, k_0}  \right ] \sigma_x 
+  \frac{k_y}{k_0}  \, \sigma_y 
\,,
\end{align}
where $ m_x $ is the quasiparticle mass along the $x$-direction, $v_y$ is the Dirac velocity along the $y$-direction, and the parameter $\delta_0$ has the dimensions of energy. We scale our Hamiltonian by $ \hbar\,v_y k_0 $ (where $k_0$ is a momentum scale) such that energy is measured in units of $ \hbar\,v_y k_0 $, mass is measured in units of $ \hbar\, k_0 v_y^{-1}$, and each momentum component is measured in units of $ k_0 $.
The system is in a (i) gapped (i.e., trivially insulating) state for $\delta_0 < 0$; (ii) gapless semi-Dirac phase at $\delta_0 = 0$; and (iii) semimetal phase for $\delta_0 > 0$ with two gapless Dirac nodes located at $\mathbf{k}_{(0)} \equiv (\pm \sqrt{2\,m_x \,\delta_0}
/\hbar , \,0)$.
We set $m_x = 0.5\,\hbar\, k_0 v_y^{-1} $ in all our computations.

\begin{figure}[] 
	\centering
\subfigure[]{\includegraphics[width=0.6\columnwidth]{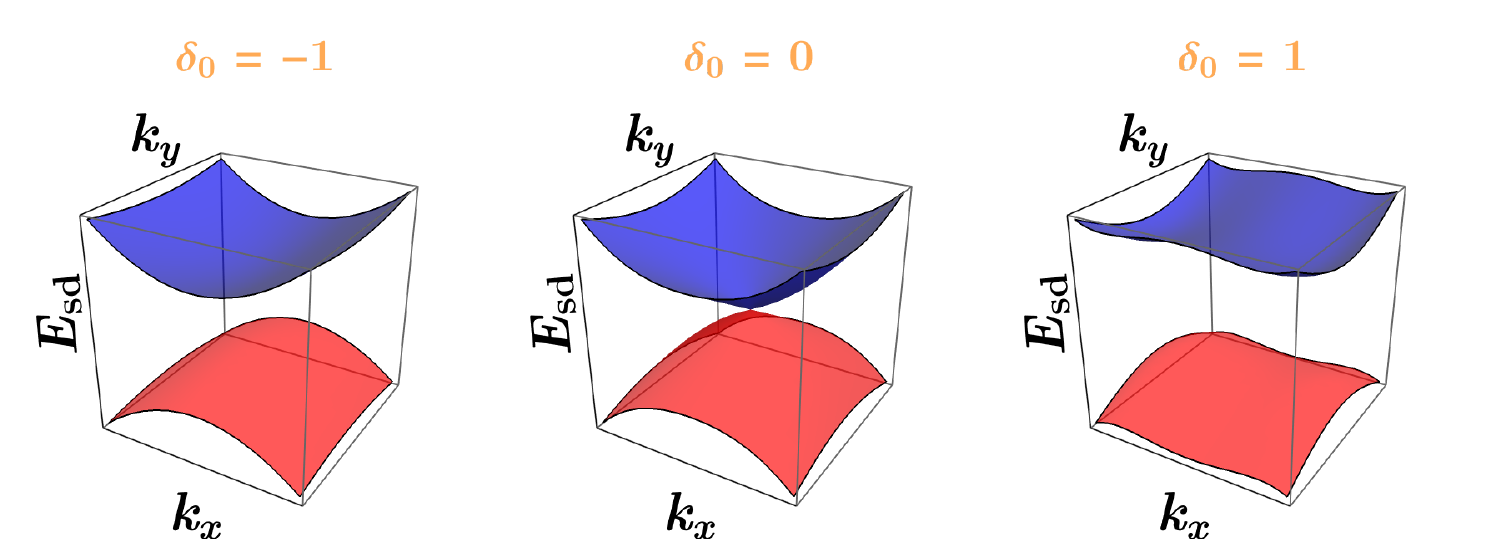}}\\
\subfigure[]{\includegraphics[width=0.6\columnwidth]{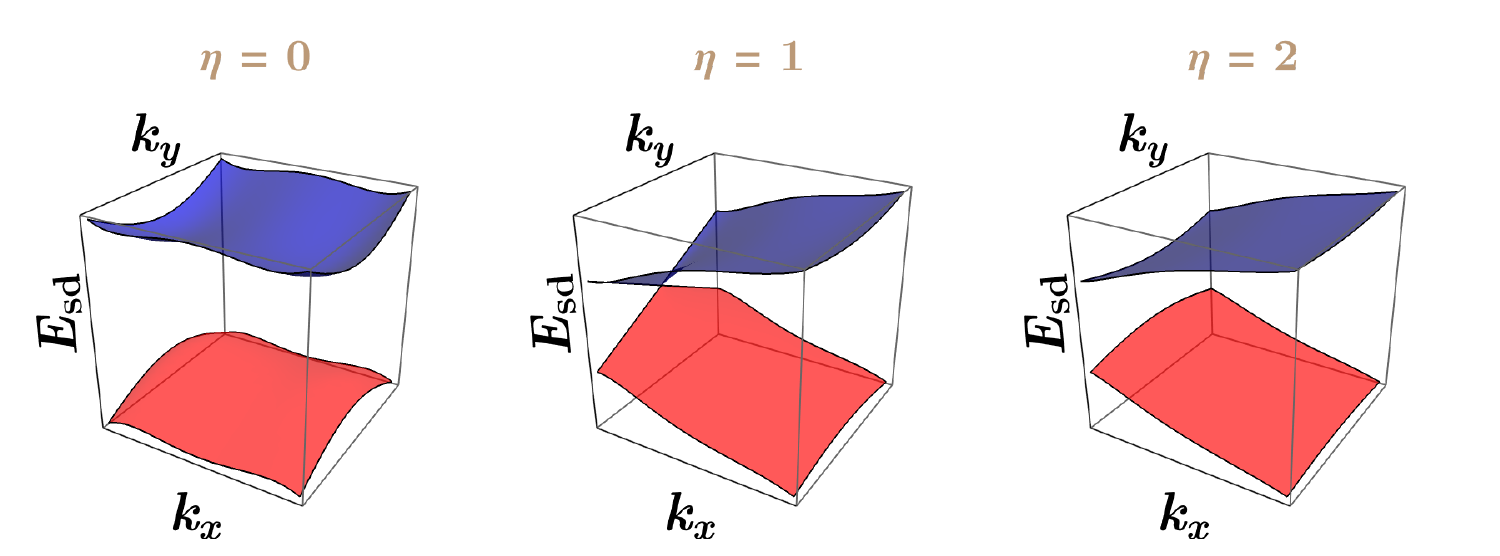}}\\
\subfigure[]{\includegraphics[width=0.4\columnwidth]{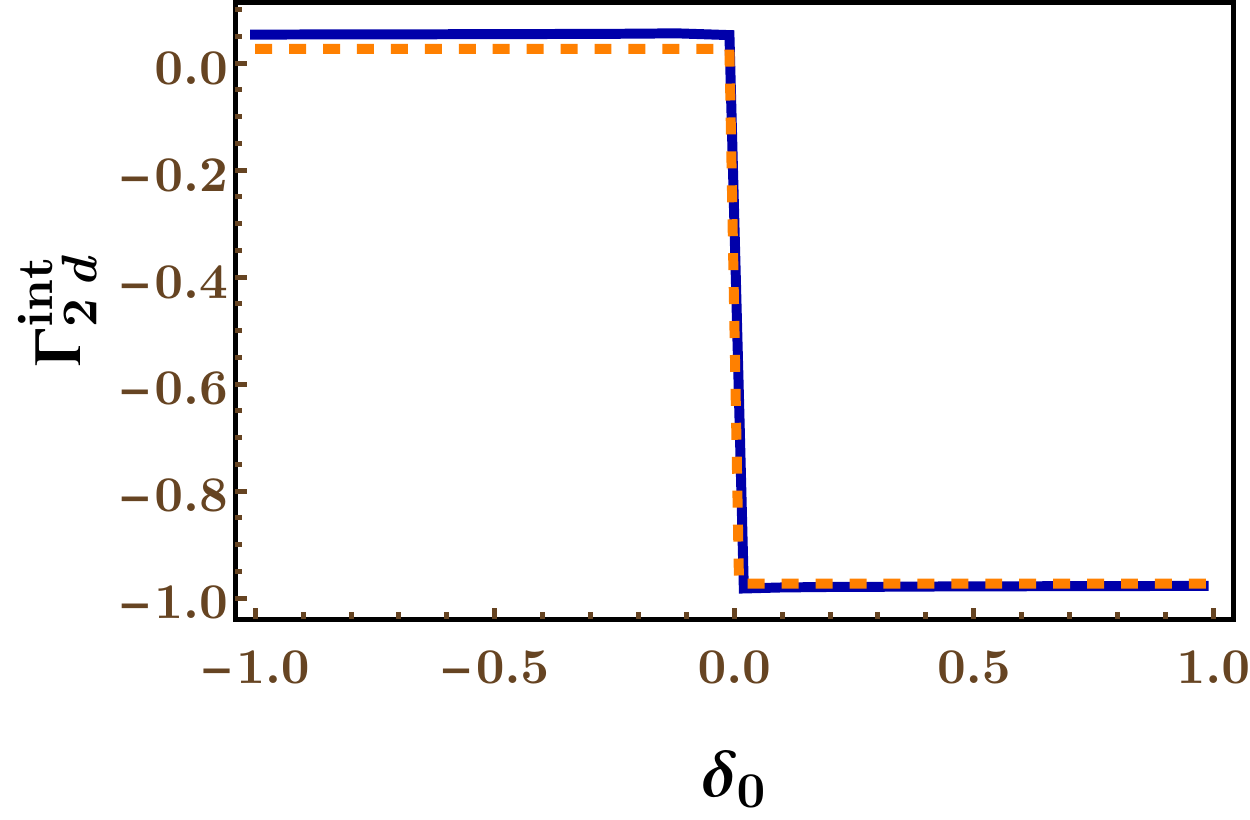}} \hspace{ 1 cm}
\subfigure[]{\includegraphics[width=0.4\columnwidth]{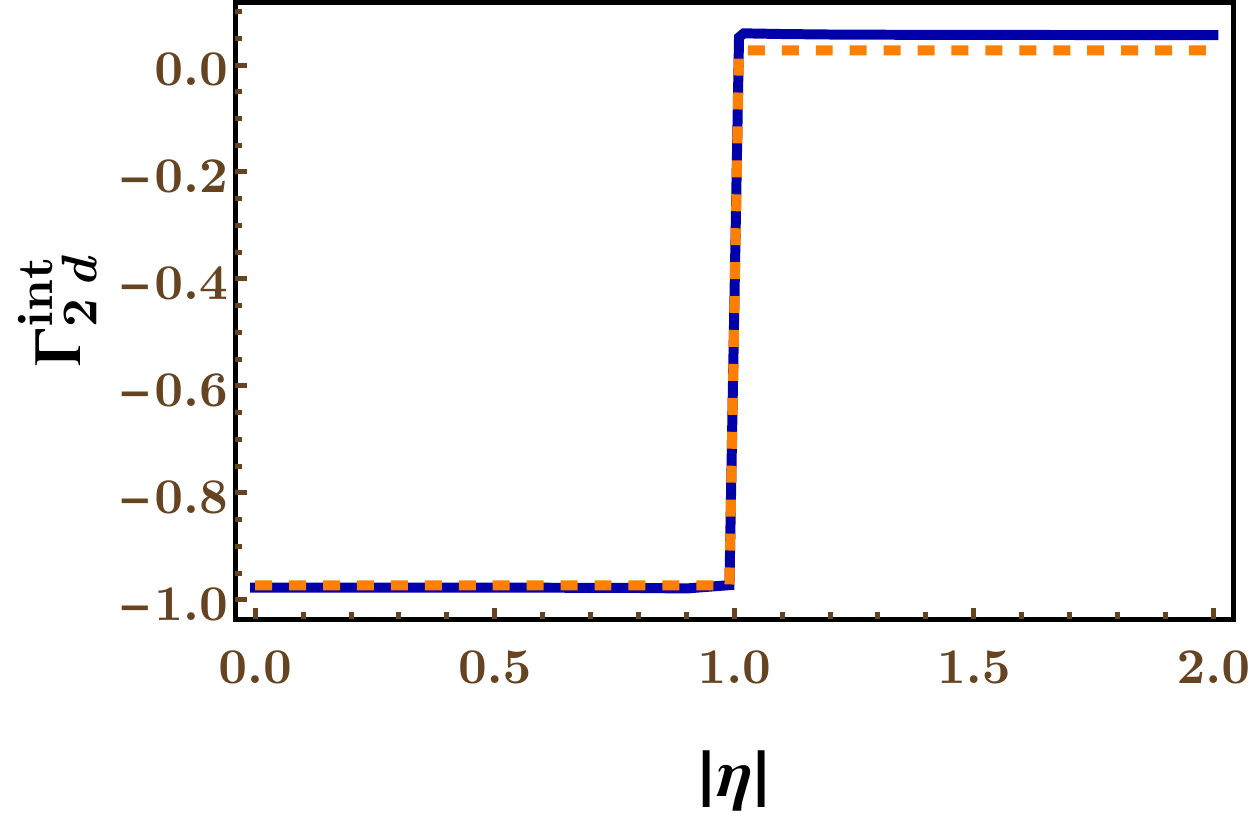}} 
\caption{
\label{fig:dir_sd}
The first two horizontal panels show the evolution of the dispersion for a semi-Dirac semimetal for $m_x = \hbar\, k_0 v_y^{-1}$ and $\beta= v_y$, in the presence of symmetry-breaking terms [cf. Eq.~\eqref{eqfullham}], as we change the various parameters in the system. In subfigure (a), the particle-hole symmetry-breaking term is absent, as $\eta $ is set to zero. In subfigure (b), the switching between distinct topological phases (on the two sides of the gap-closing point at
$\eta = \beta \, \sqrt{2\,m_x \,\delta_0 }  = 1$)
is shown by setting $\delta_0 =  \hbar\, k_0 v_y$, as $\eta$ is tuned to different values (in units of $ \hbar\, k_0 v_y $). 
In the lowermost panel, the solid blue line shows the behaviour of the 2d DIR $\Gamma^{\text{int}}_{\text{2d}} $ (in units of $\mathcal{E}^2/ \hbar^2$) for $m_x = 
\hbar\, k_0 v_y^{-1}$ and $\beta= v_y$. $\Gamma^{\text{int}}_{\text{2d}} $ is plotted as a function of
(c) $\delta_0 $ (in units of $ \hbar\, k_0 v_y$), with $ \eta = 0 $; and (d) $ |\eta| $ (in units of $ \hbar\, k_0 v_y$), with $\delta_0= \hbar\, k_0 v_y $.
We also compare its profile with $ {\mathcal C}_0$, whose value is shown using dashed orange lines.
}
\end{figure}

To $\mathcal{H}_{\text{sd}}^{0} (\*k)$, we add momentum-dependent symmetry-breaking terms of the form~\cite{saha_awf}
\begin{align}
\mathcal{H}_{\text{pert}}  (\*k)  =  
\left (  \eta + \hbar \,\beta \,k_x \right ) \sigma_z 
\Rightarrow
\frac{ \mathcal{H}_{\text{pert}}  (\*k)}   {\hbar\,v_y k_0} = 
\left (  \frac{\eta} {\hbar\,v_y k_0} 
+ \frac{\beta\, \,k_x} {v_y \,k_0 }  \right ) \sigma_z \,,
\end{align}
where $\eta $ has the dimensions of energy and $\beta$ has the units of velocity.
With this additional term, the full Hamiltonian is
\begin{align}
\label{eqfullham}
\mathcal{H}_{\text{sd}} (\*k)
= \mathcal{H}_{\text{sd}}^{0} (\*k)+ \mathcal{H}_{\text{pert}}  (\*k) 
\text{ with eigenvalues }
E_{\text{sd}}(\*k) = \pm \sqrt{\left( \frac{ \hbar^2 k_x^2} {2 \,m_x}
-\delta _0 \right)^2
+   v_y^2 \,\hbar^2 k_y^2 + \left(\eta +\beta  \,\hbar \,k_x \right)^2} \, .
\end{align}  
The additional terms $ \eta \, \sigma_z$ and $ \hbar \,\beta \, k_x\, \sigma_z$ break the particle-hole and time-reversal symmetries, respectively. In the $\delta_0 <0$, the system remains in the gapped state, which is a trivial insulator (i.e., with zero Chern number). However, the $\delta_0 > 0$ regime hosts topological insulators with different Chern numbers, as the gap closes
at $\eta = \pm \,\beta \, \sqrt{2\, \delta _0 \,m_x}$, on tuning the value of $ \eta$.
In particular, $ | \eta | < \beta \, \sqrt{2\,m_x \,\delta_0 }$ gives a nontrivial topological phase with ${\mathcal C}_0 =-1 $, whereas $ | \eta | > \beta \,\sqrt{2\,m_x \,\delta_0}$ corresponds to a trivial phase with ${\mathcal C}_0 = 0 $.
These features are demonstrated in Fig.~\ref{fig:dir_sd}. Henceforth, we set $\beta=1$ for the sake of simplicity.

From Eq.~\eqref{eqfullham}, ${\mathcal C}_0$ can be obtained directly from the Berry curvature of the LBB, which reads
\begin{align}
\Omega_{xy}(\mathbf k) = 
-\frac{
\frac{\delta _0} {k_0\, \hbar \, v_y}+\frac{k_x^2}{k_0^2}
+ 
\frac{2 \,\eta } { k_0\, \hbar \, v_y} \,\frac{k_x} {k_0}
}
{2 \,k_0^2 
\left[\left(\frac{k_x^2}{k_0^2}-\frac{\delta _0} {k_0\, \hbar \, v_y}\right)^2
+\left(\frac{\eta } {k_0 \,\hbar \,  v_y}
+\frac{k_x}{k_0}\right)^2
+\frac{k_y^2} {k_0^2}
\right]^{3/2}} \,,
\end{align}
for $m_x= 0.5\,\hbar\, k_0 v_y^{-1} $ and $\beta=v_y$.
The 2d DIR, computed numerically using Eq.~\eqref{eqdir1}, is plotted in Fig.~\ref{fig:dir_sd}. It is seen to coincide with the value of ${\mathcal C}_0$, as expected. A momentum cutoff value of $100\,k_0 $ has been used to obtain converging values in the integrals.

\subsection{Multi-Weyl semimetals} 
\label{sec:mweyl}

In this subsection, we compute the 3d DIR for 3d nodal-point semimetals.
We choose to focus on multi-Weyl semimetals~\cite{bernevig_mwsm} to understand the role of anisotropy in CD responses, as these two-band systems feature a mix of linear and higher-order dispersions, depending on the direction. They are generalizations of the Weyl semimetals (WSMs), featuring linear-in-momentum isotropic dispersions, to higher-order-in-momentum dispersions in certain directions. For the sake of comparison, we also investigate the corresponding behaviour of their isotropic counterparts, which have the same higher-order dispersion in all directions.

The low-energy continuum model of a multi-WSM is given by
\begin{align}
\label{eqhammulti}
	\mathcal{H}_{\text{aniso}}(\*k) & = 
\hbar \,v_\perp
k_0 \, \left[    \left(  \frac{k_{-}}{k_0}  \right )^J  \sigma_{+}  +
	 \left(  \frac{k_{+}}{k_0}  \right )^J   \sigma_{-} \right ]
 + \hbar \, v_z\,k_z \, \sigma_z  
 \nonumber \\
 \Rightarrow \frac{\mathcal{H}_{\text{aniso}}(\*k)} 
 {\hbar \, v_\perp  k_0 } & = 
    \left(  \frac{k_{-}}{k_0}  \right )^J  \sigma_{+}  +
	 \left(  \frac{k_{+}}{k_0}  \right )^J   \sigma_{-} 
 + \frac{v_z\, \,k_z}  {v_\perp k_0  } \, \sigma_z   \text{ with }	J \in \lbrace 1,2,3\rbrace  \,,
\end{align}
where $k_{\pm} = k_x \pm i \,k_y$ and $\sigma_{\pm} = (\sigma_x \pm i \,\sigma_y)/2$. The velocities $v_\perp$ and $v_z$ describe the Fermi velocities along the $xy$-plane and along the $z$-axis, respectively, and $ k_0$ is a material-dependent parameter with the dimensions of momentum. We set $v_z =v_\perp$ for the sake of simplicity.
The topological charge of a positive chirality multi-Weyl node is described by the integer $ J$, with the $J=1$ case simply capturing the WSM. The eigenvalues of the Hamiltonian are given by $E_{\text{aniso}}(\*k) = \pm \,\hbar\,v_\perp  k_0\, \sqrt{
\left (\frac{k_x^2+k_y^2} {k_0} \right )^J+ 
\frac{v_z^2 \, k_z^2}  {v_\perp^2  k_0^2} }$, revealing quadratic and cubic dispersions in the $xy$-plane, for a double-WSM (with $J=2$) and a triple-WSM (with $J=3$), respectively. We note that a WSM-like linear dispersion persists along the $z$-direction. Several first-principle calculations predict the existence of double-WSM in $\mathrm{SrSi_2}$~\cite{singh_srsi2}, while certain transition-metal monochalcogenides have been proposed as examples of triple-WSM behaviour~\cite{liu_triplewsm}.

\begin{figure}[] 
\centering
\subfigure[]{\includegraphics[width=0.65\columnwidth]{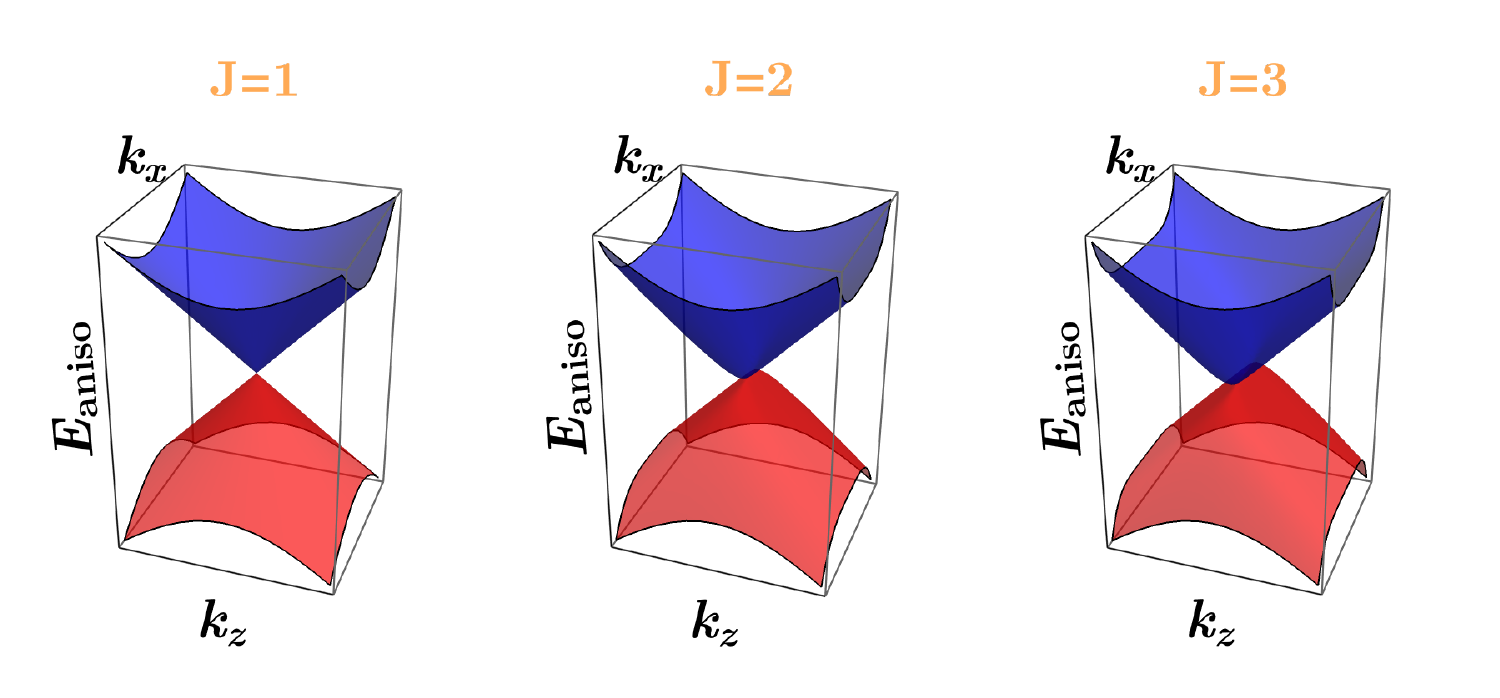}}\\
\subfigure[]{\includegraphics[width=0.4\columnwidth]{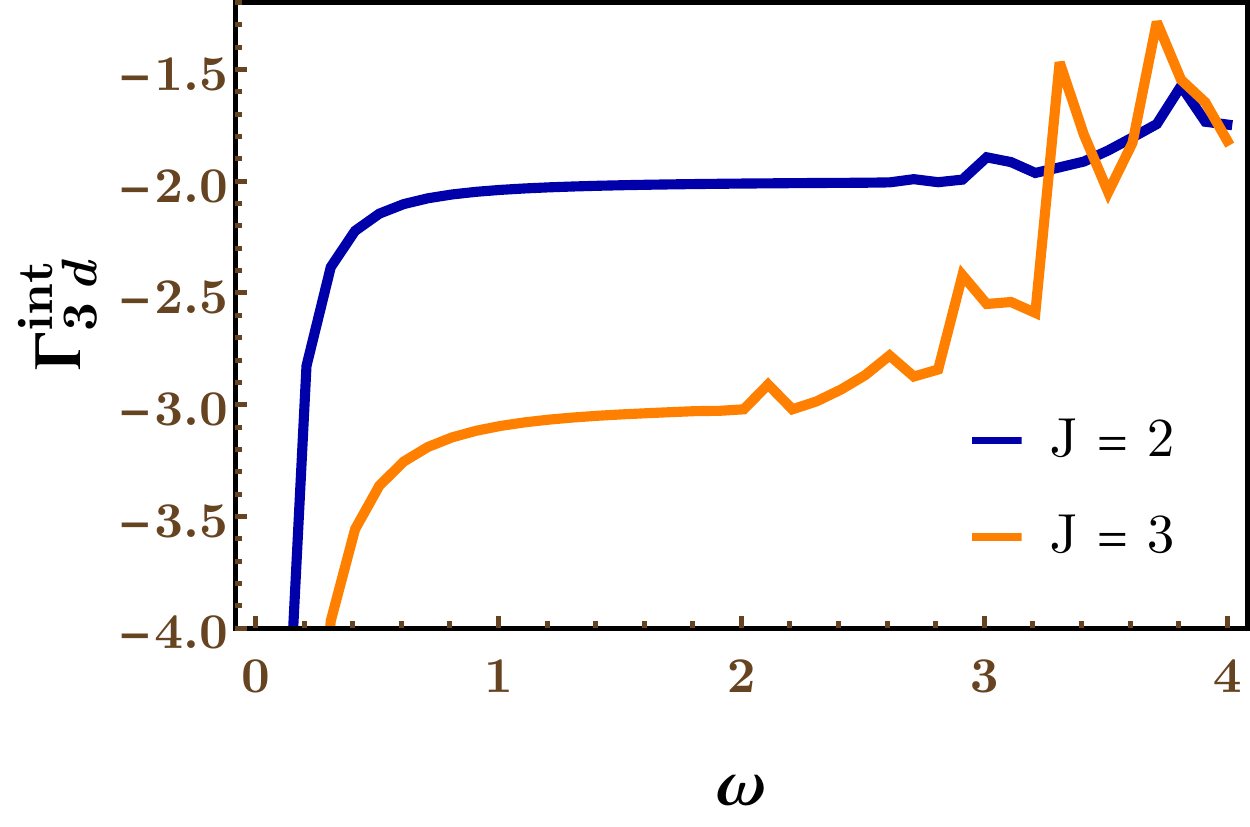}} \hspace{ 1 cm}
\subfigure[]{\includegraphics[width=0.4\columnwidth]{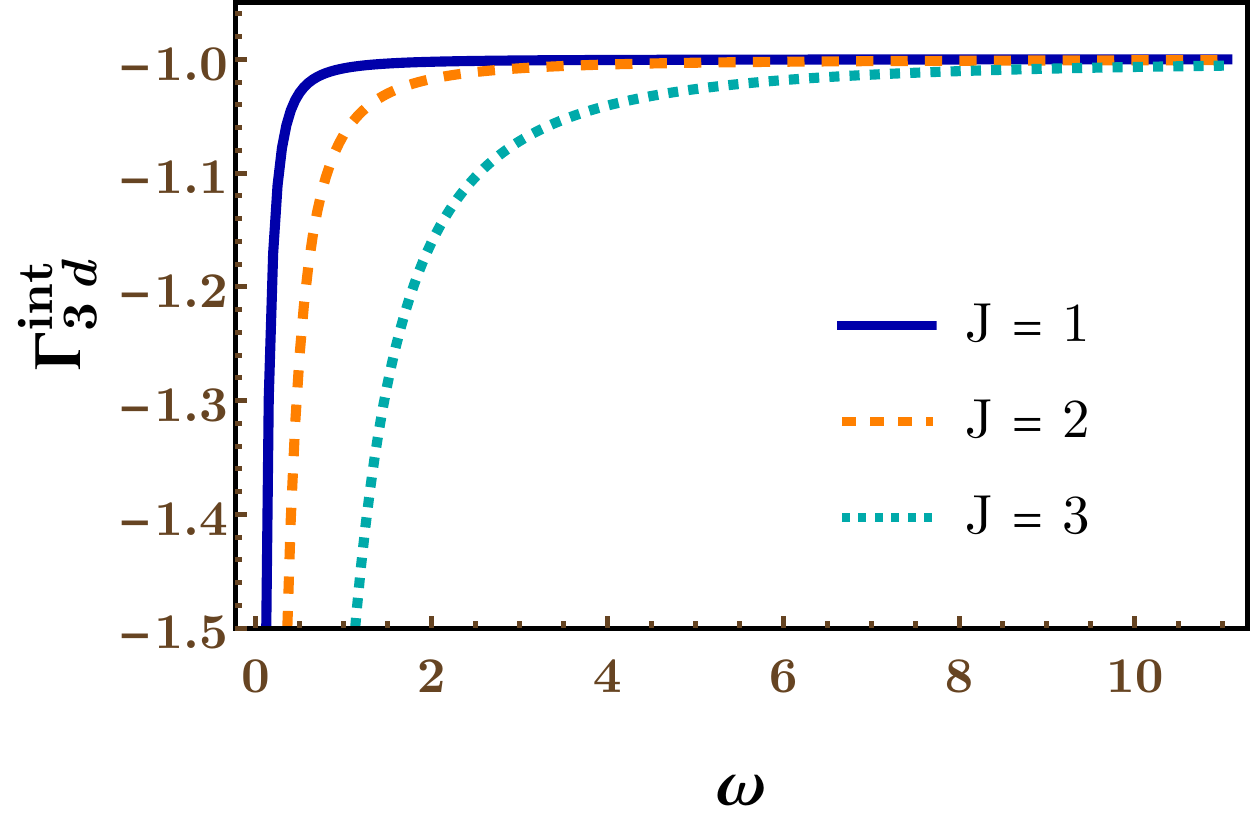}} 
\caption{
\label{fig:dir_mweyl}
Subfigure (a) shows the dispersions, obtained from Eq.~\eqref{eqhammulti}, in the $k_y=0$ plane. 
The behaviour of the 3d DIR $\Gamma^{\text{int}}_{\text{3d}} $ [in units of $\mathcal{E}^2/(2\pi \hbar^2)$] is shown as a function of the driving frequency $\omega$ (in units of $ v_\perp k_0  $) for (b) multi-Weyl semimetals given by Eq.~\eqref{eqhammulti}; (c) isotropic dispersions (where the dependence on $|\mathbf k|$ is nonlinear for $J>1$) given by Eq.~\eqref{eqiso}.}
\end{figure}

We scale our Hamiltonian by $\hbar\,v_\perp  k_0 $ such that mass and energy are measured in units of $v_\perp  k_0 $, and each momentum component is measured in units of $k_0 $.
We compare the 3d DIR results [computed using Eq.~\eqref{eqdir3d}] with the values of $\mathcal C_0$ (computed from the BC), noting that 
\begin{align}
\left \lbrace \Omega^{xy},\,\Omega^{yz},\,\Omega^{zx}
\right \rbrace
= -\frac{J \,v_z
\left (\frac{k_x^2+k_y^2}  {k_0^2} \right)^{J-1}}
{2 \,k_0^3\,v_\perp \left [ 
\left (\frac{k_x^2+k_y^2} {k_0} \right )^J
+ \frac{v_z^2 \,k_z^2} {v_\perp^2\,k_0^2} \right ]^{3/2}}
\left\{J k_z,k_x,k_y\right\} .
\end{align}
This gives $\mathcal C_0 = -J$. The 3d DIR response for $J>1$ is shown as a function of drive frequency in Fig.~\ref{fig:dir_mweyl}(a). A momentum cutoff value of $10\,k_0 $ has been used to obtain converging values in the integrals.
As expected from Eq.~\eqref{eqdir3d1}, the WSM exhibits a quantized response [cf. the $J=1$ case in Fig.~\ref{fig:dir_mweyl}(c)]. Due to anisotropy, such perfect quantization does not exist for $J>1$ throughout the frequency domain --- the response is quantized for a limited frequency range, and as the frequency increases beyond a critical value, the response decays from the quantized plateau to eventually become zero. We note that the rate of decay is faster for higher value of $J$. This can be understood if we consider the fact that quantization of DIR is highly sensitive to the magnitude of anisotropy. With increasing driving frequency, photons probe the higher and higher energy ranges, where the effects of anisotropy are greater. And these effects grow more rapidly with the increasing power of momentum in the nonlinearly dispersing directions. This, in turn, destroys the quantization faster.

We now compare the response seen for the multi-WSMs with that of the corresponding isotropic cases, which have the same higher power-law dependence on $|\mathbf k|$ as seen for the multi-WSMs on $(k_x^2+k_y^2)$ in the $xy$-plane. Hence, we consider the Hamiltonian~\cite{ahn_mweyl}
\begin{align}
\label{eqiso}
\mathcal{H}_{\text{iso}}(\*k) = \hbar\,v_\perp k_0 
 \left(  \frac{|\*k|}{k_0}  \right )^J \, \hat{\*k} \cdot \boldsymbol{\sigma} \,,
\end{align}
with energy eigenvalues $\pm \, \hbar\,v_\perp k_0  \left(   {|\*k|}/ {k_0}  \right )^J $.
Here, $ v_\perp$ and $k_0$ are again material-dependent parameters and we scale the Hamiltonian by the factor $\hbar \,v_\perp k_0 $ as before. Irrespective of the value of $J$, ${\mathcal C}_0 = -1$ for these isotropic systems, since $\Omega^{\mu \nu} 
=-\frac{\epsilon^{\mu\nu\lambda} \,k_\lambda}
{2 \,|\mathbf k|^{3}}$. The 3d DIR response is shown as a function of the drive frequency in Fig.~\ref{fig:dir_mweyl}(c), which indeed shows saturation to the value of $-1$ in units of $\mathcal{E}^2/(2 \pi \hbar^2)$. However, we note that the value of $\omega$, at which saturation is reached, increases with increasing nonlinearity of the dispersion.

\section{Summary and outlook}
\label{sec:summary} 

In this paper, we have investigated circular dichroism for nodal-point semimetals harbouring nontrivial topological charges. The response is initiated by a circular periodic drive, which causes transitions from a filled band to higher energy bands. The circularly dichroic response is caused by unequal transition rates connected with the chiral nature of the systems.

We would like to emphasize that the 3d DIR response~\cite{sajid_cd} (applicable for 3d systems) is different from the 2d DIR formula~\cite{goldman_dir} (apllicable for 2d systems). For computing $\Gamma^{\text{int}}_{\text{3d}} (\omega)$, the net transition rate$\Delta \gamma_{mn}^{\mu\nu} $ is computed for each 2d projection of the 3d system, which is confined to the $\mu\nu$-plane. With $\lambda$ denoting the direction perpendicular to this 2d plane under consideration, we then multiply $\Delta \gamma_{mn} ^{\mu\nu}$ with the band velocity difference  $\partial_{k_\lambda } E_{mn}$, and integrate it over the entire 3d Brillouin zone to get the differential current along the $\lambda$ direction. This is followed by a summation over the three mutually perpendicular directions. This complicated expression happens to overlap with the expression of the Chern number of the LBB, at leading order, only for systems with isotropic dispersions \cite{sajid_cd}. We also note that $\Gamma^{\text{int}}_{\text{3d}} (\omega)$ is a function of $\omega$ (i.e., integration over $\omega$ is not performed while deriving it). On the other hand, $\Gamma^{int}_{\text{2d}} (\omega)$ is obtained by integrating over a frequency interval.

For 3d systems with isotropic dispersions (both linear and nonlinear), the 3d DIR exhibits a quantized response in terms of the LBB Chern number. This robust quantization persists throughout the frequency domain, and is absent for anisotropic multi-WSMS (with $J>1$), which feature quadratic (for $J=2$) and cubic (for $J=3$) dispersions in the plane perpendicular to the $z$-axis (in our choice of coordinate system). Instead, the multi-WSMs show a small quantized plateau followed by a non-quantized response, which decays to zero as $\omega$ increases. 

For the 2d semimetals, the 2d DIR response does not depend on anisotropy. The analytical expression in fact shows that it is always proportional to the Chern number of the filled band, and this is what we find in our numerical results as well.

Our CD computations complement earlier studies of DIR on various systems~\cite{goldman_dir,sajid_cd}. The DIR response also supplements other techniques, involving time-periodic drives, to characterize topological phases~\cite{nonlin-photo,ips-sandip,ips-sandip-sajid,shivam-serena}. In future, it will be worthwhile to study the CD response in the presence of interactions and/or impurities~\cite{rahul-sid,ipsita-rahul,ips-qbt-sc,ips-biref,ips-klaus}.
The presence of very strong interactions may induce many-body effects like (i) destroying the quantization of topological charges like Chern numbers~\cite{kozii,Mandal_2020}; (ii) emergence of strongly correlated phases~\cite{ips-seb,MoonXuKimBalents,rahul-sid,ipsita-rahul,ips-qbt-sc,ips-biref}, where the quasiparticle picture breaks down.

\section*{Acknowledgments}

We thank Soustav Bose and Sajid Sekh for participating in the preliminary stages of this work.

\bibliography{ref.bib}

\end{document}